\documentclass[aip, jcp, floatfix, reprint]{revtex4-1} %% preprint -> onecolumn, reprint -> twocolumn

\usepackage{amsmath, amssymb}
\usepackage{booktabs, csquotes, fancybox, graphicx, tabularx, wrapfig, xfrac}
\usepackage[dvipsnames]{xcolor}

\newcommand{\picsize}{0.48\textwidth}
\newcommand{\etal}{\textit{et\,al.} } 
\newcommand{\kb}{k_\mathrm{B}}
\newcommand{\diff}{\mathrm{d}}
\renewcommand{\exp}[1]{\mathrm{exp}\left(#1\right)} 

\newcommand{\update}[1]{\textcolor{Black}{#1}}

\begin{document}
\title{Exploring the free energy gain of phase separation via Markov State Modeling}
\date{\today}
\author{Myra Biedermann}
\email{m.biedermann@uni-muenster.de}
\author{Andreas Heuer}
\email{andheuer@uni-muenster.de}
\affiliation{Institute of Physical Chemistry, University of Münster, Corrensstraße 28/30, 48149 Münster, Germany}
\affiliation{Center of Nonlinear Science (CeNoS), University of M\"unster, Corrensstr. 2, 48149 M\"unster, Germany}
\affiliation{Center for Multiscale Theory and Computation (CMTC),  University of M\"unster, Corrensstr. 40, 48149 M\"unster, Germany}

\begin{abstract}
The gain of free energy upon unmixing is determined via application of Markov state modeling (MSM), using an Ising model with a fixed number of up- and down-spins. MSM yields reasonable estimates of the free energies. However, a closer look reveals significant differences which point to residual non-Markovian effects. \update{These non-Markovian effects are rather unexpected since the typical criteria to study the quality of Markovianity indicate complete Markovian behavior.} We identify the sparse connectivity between different Markov states as a likely reason for the observed bias. By studying a simple five state model system we can analytically elucidate different sources of the bias and thus explain the different deviations that were observed for the Ising model. Based on this insight we can modify the determination of the count matrix in the MSM approach. In this way, the estimation of the free energy is significantly improved.
\end{abstract}

\maketitle

\section{Introduction}
Studying the free energy as a function of appropriately chosen order parameters yields important insight into the thermodynamic and dynamic properties of complex systems. During the last two decades, Markov state modeling (MSM) has gained increased attention as a sampling method for these free energy landscapes\cite{Pande2009, BowmanPande2010, IntroMSM}. The development of MSM has been largely driven by studies of biomolecules and especially protein folding\cite{Chodera2014}. There, Markov state models enabled the investigation and detailed comparison with experiment of proteins which fold on the millisecond timescale\cite{Bowman2010, Voelz2010, Voelz2012}.
MSM approximates the long-time statistical dynamics of a system by a Markov chain on a discrete partition of the phase space. Instead of considering single trajectories, as it is done e.g. in classical molecular dynamics, MSM focuses on ensemble dynamics\cite{Noe2011, Schutte1999PCCA}. Essentially, MSM consists of a partitioning scheme of the system's phase space into discretized states and a transition probability matrix that contains the conditional probabilities for the system to transition from one of the discrete states to another within a certain time interval $\tau$, which is called the lag time\cite{Noe2011}. This transition probability matrix contains all necessary information to compute the system's free energy landscape, including detailed information about the thermodynamics and dynamics\cite{Pande2009}.

Finding a suitable partitioning scheme for a system's phase space that is able to capture the important slow relaxation processes in the system is in general a highly non-trivial task. Due to the high dimensionality of a typical phase space, MSM usually starts by somehow partitioning the observed configurations into microstates, followed by some kinetically relevant clustering method that further coarse-grains the model.\cite{Noe2011, Pande2010} Straightforward approaches such as the root mean square distance to a reference structure, radius of gyration or chemically intuitive order parameters such as specific distances or angles are often very inaccurate partitioning methods\cite{PerezHernandez2013, Noe2008, Brezovska2012} and thus fail to reveal the important kinetically metastable states of the system. Much effort has been made in this direction in order to be able to identify "good" order parameters or reaction coordinates\cite{McGibbon2016, PerezHernandez2013} and decompose the configuration space into kinetically metastable states\cite{Chodera2007}. For example, a method has been proposed recently that is able to automatically identify both metastable and transition state regions, which are necessary to describe the slowest process in a system due to their control over the system's overall kinetics\cite{Martini2016}.

Because only conditional transition probabilities are necessary to build a Markov state model, trajectories that are significantly shorter than the longest relaxation time of the system can be used. This implies that the trajectories only have to be long enough to reach local equilibrium within the sampled part of the configuration space instead of requiring them to achieve global equilibrium\cite{Noe2011}. Instead, MSM enables a divide and conquer approach by computing many (short) trajectories from different starting points in the system's phase space, possibly in parallel, and hereby mitigating the sampling problem that one often encounters when simulating large and complex systems\cite{Elmer2005, Chodera2006, Noe2009}.

In this work, we explore the possibility to use MSM for studying the free energy gain upon phase separation. Low temperature mixtures of different molecular species typically display a strong thermodynamic driving force for phase separation. Thus, the ability of MSM to extract information from many short simulations, starting at the mixed state, might be ideal for extracting the relevant thermodynamic information. In order to study the key features of this application, we employ a 2D Ising model with conserved number of up- and down-spins. The Ising model has been frequently used as a simple, but still highly non-trivial model to study the physical mechanisms of phase separation\cite{Binder1974, Flinn1974, Amar_etal1988, Fratzl_etal1991}. On first sight, MSM can be used to determine the free energy difference between the initially mixed and the subsequent unmixed phase. However, a closer look reveals significant non-Markovian effects, although the standard analysis, such as a check of the Chapman-Kolmogorov relation, suggests perfectly Markovian behavior. Via some model analysis we can trace this effect back to generic properties of phase-separating systems. This insight allows us to improve the MSM algorithm. 

The outline of the paper is as follows. In Sect. \ref{sec:methods} we start with an introduction of the model and the numerical analysis. Subsequently, MSM is applied in Sect. \ref{sec:MSMresults}, including positive checks for Markovian behavior. This allows us to determine the change in free energy upon unmixing. Comparison with the results of an alternative free energy method reveals significant deviations which point towards the presence of subtle non-Markovian effects that were not reflected by the results of the Markov validation methods. Thus, we refine the discretization scheme and scrutinize the connectivity of the different microstates in Sect. \ref{sec:SubdivisionNstates}. Then, in Sect. \ref{sec:5statemodel}, we discuss a minimalistic five state model which is defined such that it also contains the observed connectivity effects. For this model we can fully understand the origin of non-Markovianity in the MSM estimation process. In Sect. \ref{sec:AbsentConnectionsAlgorithm} we extend the MSM algorithm based on the insights from the previous Section. In this way a significant improvement of the estimation is achieved. We conclude with a discussion and a summary in Sect. \ref{sec:conclusion}.

\section{Methods} \label{sec:methods}

\subsection{The model system}\label{sec:methods-TheModelSystem}
The Ising model consists of spins on a lattice that can only exhibit two discrete states: spin up or spin down ($s = \pm 1$)\cite{Ising1924}.
The Hamiltonian, $H$, of the Ising system is given by
	\begin{align}
		H / T = - \sum\limits_{\langle ij\rangle} J_{ij} s_i s_j, \label{eq:Hamiltonian}
	\end{align}
where $\langle ij\rangle$ denotes that the sum goes over all pairs of adjacent (or neighboring) spins on the lattice, $J_{ij}$ is the interaction strength parameter between spins $i$ and $j$ and $T$ is the temperature. On a two-dimensional lattice each spin has four nearest neighbors. We set the Boltzmann constant $\kb$ to $1$ in all computations of physical properties such that all thermodynamic observables and also the interaction parameter $J$ become dimensionless.
	
We aim at studying the process of phase separation in a binary mixture. To do so, we fix the concentration of up- and down-spins during a simulation and set $J_{ij}=0$ for spins of identical type ($s_i = s_j$), so that only neighboring spins of opposite type contribute to the Hamiltonian of the system ($J_{ij}=J$ for $s_i \neq s_j$). We use square lattices with side length $m$ and a 1:1 ratio of both spin types. Hence, the total number of spins $n=m^2$. Periodic boundary conditions are applied to both dimensions in all simulations in order to avoid boundary effects\cite{FrenkelSmit}.

We employ a Metropolis Monte Carlo algorithm\cite{Metropolis1953} to generate trajectories of this Ising system. Instead of flipping spin states, as it is done in the original Ising model, one Monte Carlo (MC) step in our system consists of randomly picking two spins on the lattice and attempting to exchange the positions of these spins according to the Metropolis criterion. These two spins can be located anywhere on the lattice and do not necessarily have to be nearest neighbors. By doing so, we ensure that the ratio of up-spins to down-spins is maintained during a simulation. Initial configurations are generated by randomly distributing a 1:1 ratio of up-spins to down-spins on the lattice.
		
The total number of possible configurations in these systems can be computed using the binomial coefficient.
%		\begin{align}
%			\Omega = \binom{n}{n_{+}} = \frac{n!}{n_{+}! (n-n_{+})!},
%		\end{align}						
%where $\Omega$ denotes the number of possible configurations, $n$ is the total number of spins (with $n=n_{+} + n_{-}$) and $n_{\rm +}$ is the number of up-spins. 
Already for a very small lattice with $n=m^2=6^2=36$, the total number of possible configurations is $9\, 075\, 135\, 300$. Thus, grouping similar configurations together in order to reduce the number of distinct states during the computation of thermodynamic observables is not only desirable but also obligatory if one intents to use Markov state modeling to estimate these observables. \update{Using the nearest-neighbors as an indicator of the degree of phase-separation has been successful in studies of more complex, molecular phase-separating systems\cite{Hakobyan2013} and thus seems to be a reasonable order parameter also for the phase separating Ising system. Here, this translates to counting the number of spins of opposite type sitting next to each other on the lattice (also referred to as "$+-$ pairs" later on).}
%One possible and straightforward way to do so can be deduced from the Hamiltonian of the system, eq. \eqref{eq:Hamiltonian}. The interaction parameter is only non-zero for pairs of spins of opposite type and the sum includes only neighboring spins on the lattice. Thus, it becomes clear that the only contributions to the Hamiltonian result from spins of opposite type sitting next to each other on the lattice (also referred to as "$+-$ pairs" later on). In consequence, if the total number of those pairs on the lattice is known, the Hamiltonian of a configuration is completely determined independent of the actual positions of these pairs on the lattice. Hence, the generic option is to group all those configurations together that have the same number of $+-$ pairs and thus the same energy. 
All possible numbers of $+-$ pairs are even numbers. Thus, for reasons of convenience, the order parameter $N$ chosen in this work to characterize the configurations is exactly one half of the total number of $+-$ pairs in a configuration. \update{Since the interaction parameter between spins in our systems is only non-zero for pairs of spins of opposite type, it becomes clear that all configurations with equal numbers of $+-$ pairs will also have the same energy.} Introducing this order parameter to the Hamiltonian results in
		\begin{align}
			H / T = J \, 2 N.
		\end{align}		
\update{As a consequence the order parameter fully determines the energy of the Ising model. For general phase-separating systems one expects a significant but not perfect correlation of the order parameter and the potential energy of that system. Please note that the results, discussed in this work, do not depend on the interchangeability of energy and order parameter.}

\begin{figure}
	\includegraphics[width=\picsize]{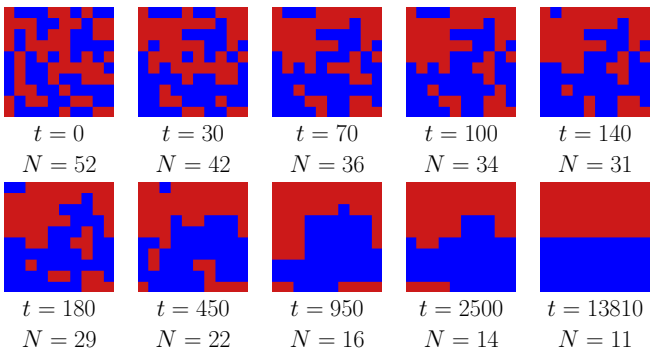}
	\caption{Snapshots of a MC simulation of a system of size $m^2=10^2$ with $J=2.5$ that illustrate the evolution of the system from a randomly mixed configuration towards a phase-separated configuration. The images are captioned with their respective number of MC steps, $t$, and their corresponding order parameter, $N$.}
	\label{fig:trajectory_snapshots}
\end{figure}

Fig. \ref{fig:trajectory_snapshots} displays snapshots of a MC simulation of a $10^2$ sized system with $J=2.5$. Here, the starting configuration is a mixed configuration with order parameter $N=52$, which corresponds to 104 $+-$ pairs on the lattice. The snapshots demonstrate the general behavior of the Ising system during a MC simulation: because the energy of the system increases with increasing number of $+-$ pairs, this drives the system to evolve towards configurations with less $+-$ pairs, which correspond to phase-separated configurations. Furthermore, Fig. \ref{fig:trajectory_snapshots} demonstrates how the order parameter indicates the degree of phase separation: high values of $N$ correspond to mixed configurations, whereas decreasing values of $N$ indicate an increase in phase separation.

For reasons of convenience, we introduce one modification to the definition of the order parameter: there is always one value of $N$ next to the minimal possible value and two values in the high-$N$ limit which topologically cannot be realized. In order to obtain a continuous range of order parameter values, we map the lowest possible order parameter to its next higher value, $N_\mathrm{min}\rightarrow N_\mathrm{min}+1$, s.t. the perfectly unmixed state in a $6^2$ sized system corresponds to $N=7$ although it really has $2\times 6$ $+-$ pairs. For the same reason, we also transform the highest possible order parameter value via $N_\mathrm{max} \rightarrow N_\mathrm{max}-2$. This mapping procedure is only used for identifying the states and not during energy computations.

\subsection{Analysis methods}\label{sec:methods-SimDetails}
We use MC simulations with randomly chosen mixed configurations as initial starting points (see also section \ref{sec:methods-TheModelSystem}) to generate sets of simulation data. In order to achieve good sampling in both the phase-separated region and the mixed region of the configuration space, our data sets consist of both long trajectories and many short trajectories.  Long trajectories contain $5\, 000$ MC steps for the $6^2$ sized systems, $20\, 000$ MC steps for the $10^2$ systems, $50\, 000$ MC steps for the $14^2$ systems and $60\, 000$ MC steps for the $20^2$ systems, respectively, and the respective number is divided by ten for the short trajectories in each case. Every data set consists of $500$ long trajectories and $25\, 000$ short trajectories. We aim at minimizing the influence of statistical errors to our MSM estimates, which we survey by computing the average and standard deviation of the stationary distribution from ten independent Markov models that have been build from independent data sets. By doing so, we ensure that the results presented in the following sections are basically void of effects of statistical uncertainty.

To build a Markov state model from a set of simulation data, the MC trajectories are projected onto the discretized order parameter states by assigning each configuration that occurs in a trajectory to its corresponding order parameter. Then, transitions between these order parameter states at lag time $\tau$ are harvested by going through the trajectories in a sliding window mode\cite{Noe2011}, i.e. all transitions from time pairs $(1\rightarrow\tau), (2\rightarrow\tau+1), ...$ are considered. This results in a $(l\times l)$ transition count matrix $\mathbf{C}$, where $l$ is the total number of order parameter states and $C_{ij}$ contains the total number of transitions from state $i$ to state $j$ that occur in the simulation data. From this count matrix we compute the reversible maximum likelihood estimate for the transition probability matrix, $\mathbf{\hat{T}}$, and the corresponding stationary distribution vector, $\hat{\boldsymbol\pi}$, in an iterative fashion as proposed in refs. \citenum{Pande2009} and \citenum{Noe2015}. Here and hereafter, the hat denotes an estimator. We terminate the iteration when the norm of the change of $\hat{\boldsymbol\pi}$ in one iteration step is smaller than $10^{-10}$.

Subsequent to the estimation process, we compute the implied timescales\cite{Swope2004, Swope2004-applications} of the estimated model and perform Chapman-Kolmogorov tests\cite{Noe2011} in order to validate the estimated Markov state model.

Once that a Markov state model has been estimated and validated, we can derive thermodynamic properties of the system from the model. The free energies of the order parameter states can be computed from the equilibrium probabilities, apart from a constant, according to
\begin{align}
	F(N)/T = -\ln\left( \pi(N) \right). \label{eq:freeEnergy}
\end{align}
and the entropy of the system is given by the logarithm of the density of states, $g(N) = \exp{S(N)}$, in the Boltzmann equation,
\begin{align}
	\pi(N) = g(N)\,\exp{-H(N)/T}, \label{eq:Boltzmann}
\end{align}
where we have again set the Boltzmann constant $\kb=1$.
It should be noted that the density of states in the Ising system is not an actual density but rather refers to the number of configurations that correspond to a given energy, i.e. to a given order parameter. Therefore, when we employ the term "density of states" hereafter, we actually mean the number of states for a given energy.

In order to compare the result from an estimated Markov model to results from an alternative sampling method, we employ the Wang-Landau (WL) sampling method\cite{WangLandau2001PRL, WangLandau2001PRE, Landau2004}. The WL algorithm computes the density of states of a system by performing a Monte Carlo like random walk in energy space. As proposed in ref. \citenum{Landau2004}, an initial value of the modification parameter of $\ln(f_\mathrm{initial})=1$, a final value of $\ln(f_\mathrm{final})=10^{-8}$ and a flatness criterion of at least 0.85 is used here ($0.85$ for system size $20^2$, $0.9$ for system sizes $14^2$ and $10^2$, $0.975$ for system size $6^2$). We are aware of recent work considering the saturation of the error in Wang-Landau sampling and the resulting non-convergence of this method\cite{Belardinelli2007JChemPhys}. However, when comparing the WL results for this Ising system with results obtained via the alternative $1/t$ algorithm\cite{Belardinelli2007PhysRevE}, we observe good agreement and therefore continue to use WL sampling as a reference method.

It should be noted that the density of states that is computed via \eqref{eq:Boltzmann} from MSM or via the WL algorithm is only a relative density. We choose to normalize our results by shifting the curves such that the density of states of the minimal possible order parameter is set to 1. There is only one distinct type of configuration (and its translational shifted or rotated analogues) which corresponds to the minimal possible order parameter and it consists of the perfectly phase-separated configuration with two blocks of up- and down-spins, respectively, that share a smooth boundary (see the configuration at $13810$\,MC~steps in Fig. \ref{fig:trajectory_snapshots} for a visualization).

All calculations for the five state model system in Sect. \ref{sec:5statemodel} have been performed with the software Mathematica\cite{mathematica}.

\section{MSM analysis}\label{sec:MSMresults}

\subsection{MSM Validation}
In this part, we focus on Markov state models of very small Ising systems with size $m^2=6^2$ in order to demonstrate the validation methods. We used the same validation methods for larger systems and observed analogous results but omit to show the corresponding data for the sake of brevity.

A Markov state model assumes that transitions between the discretized (Markov) states of the system obey the Markov property, i.e. that these jumps between states are memoryless. Apart from the influence of statistical uncertainties that arise due to finite sampling, the quality of a Markov model depends on the discretization scheme that is used to partition the phase space of the system into Markov states and the lag time at which the Markov model is estimated. A Markov model cannot resolve the dynamics of a system within a discretized state and thus assumes instant local equilibration within this state\cite{Noe2011}. Therefore, both the discretization scheme and the lag time should to be chosen such that this assumption of Markovian jumps between discrete states is approximately fulfilled. In general, the finer the discretization and the longer the lag time, the better the resulting Markov model\cite{Noe2011}. We employ two validation methods that have been proposed in the literature to test the quality of our Markov state models: the implied timescales\cite{Swope2004, Swope2004-applications} and the Chapman-Kolmogorov test\cite{Noe2011}.

The implied timescales (ITS) refer to the physical relaxation times of a system. Their values can be obtained from the eigenvalues $\lambda_i$ of the estimated transition probability matrix: $t_i = -\frac{\tau}{\ln \lambda_i}$. For a good Markov model with sufficient sampling and negligible statistical errors there should be a range of lag times $\tau$ in which the corresponding slowest implied timescales are approximately constant because the ITS are physical properties of the system and therefore should be independent of the lag time at which the model has been estimated\cite{Swope2004}. Although observing constancy in the implied timescales is not a strict test of Markovianity, it has been empirically observed that constant implied timescales strongly indicate that the Markov model approximates the underlying dynamics well\cite{Chodera2006}.

\begin{figure}
	\includegraphics[width=\picsize]{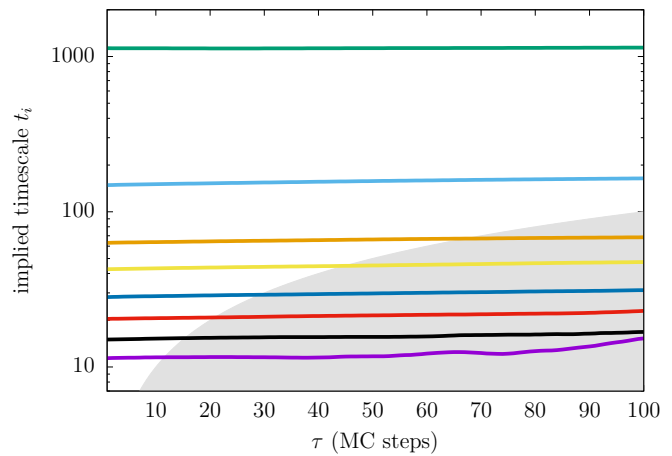}
	\caption{The nine largest implied timescales in an Ising system with $J=1.5$ plotted logarithmically as a function of the lag time $\tau$. The grey area represents the region where the estimates for the timescales are not reliable because $\tau>t_i$ and so the process under investigation has already decayed.}
	\label{fig:ITS6}
\end{figure}

Fig. \ref{fig:ITS6} displays the nine largest implied timescales for a system with interaction parameter $J=1.5$. The grey area represents the region where the lag time is larger than the timescale and so the process under investigation has already decayed during the time interval of one lag time. Estimates for the ITS in this region are therefore not reliable. Outside of this area, all nine displayed ITS show approximately constant behavior over the whole range of lag times $0<\tau\leq 100\,$MC~steps that is shown. We observed similar approximate constancy for different interaction parameters. This indicates that a lag time of a few MC steps seems to be sufficient in order to obtain approximately Markovian transitions between the order parameter states.

However, the implied timescales are not a strict test of Markovianity because although true Markovian dynamics imply constancy of implied timescales in $\tau$, the reverse statement is not true because this would also require the eigenvectors to be constant\cite{Noe2011}. For this, a Chapman-Kolmogorov (CK) test is necessary, which tests whether lag time and discretization have been chosen such that the MSM predictions for the evolution of the system are consistent with the simulation data. We employ the implementation proposed in ref. \citenum{Noe2011}. There, the transition probability to jump from a certain state $i$ to a state $j$ within a time interval $k\tau$ is computed both from the Markov model that has been estimated at lag time $\tau$ as well as from the (pure) simulation data and then plotted as a function of $k\tau$. To account for statistical uncertainties that arise due to finite sampling, the transition probabilities from the simulation data are plotted with their one-sigma standard error.

\begin{figure}
	\includegraphics[width=\picsize]{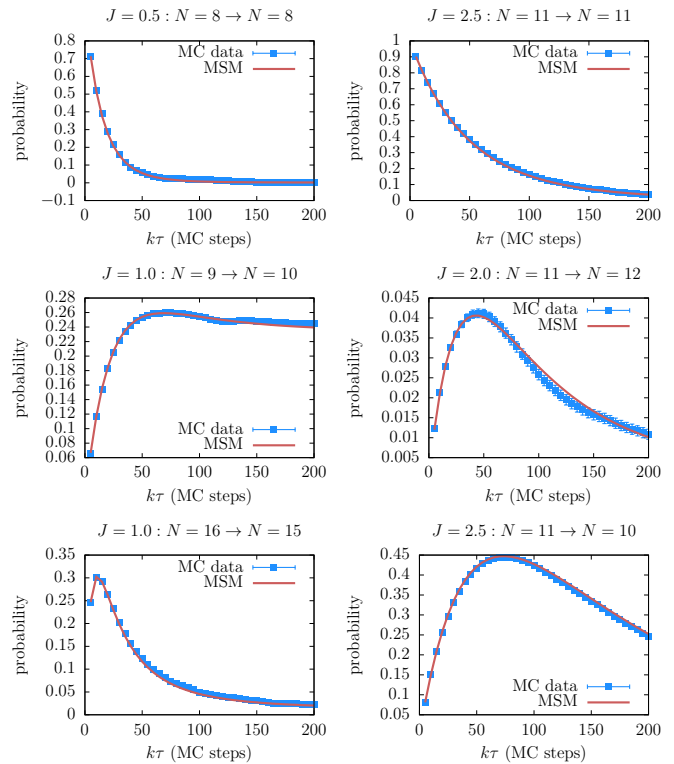}
	\caption{Plots from a Chapman-Kolmogorov test where the probabilities for transitions between Markov states that are predicted by a Markov model (red curve) are compared to the probabilities given by the simulation data (blue symbols). The corresponding Markov models were estimated at $\tau=5$\,MC~steps.}
	\label{fig:CKtest}
\end{figure}

Fig. \ref{fig:CKtest} displays some example transition probabilities between order parameter states from systems with varying interaction parameter. The corresponding Markov state models have been estimated at lag time $\tau=5$\,MC~steps. We focus on the probabilities to stay in a certain state within the given time interval $k\tau$, with $k \in \mathbb{N}$, and the probabilities for jumps between neighboring states in order to avoid comparing very small numbers. The agreement between the probabilities that are predicted by MSM and the probabilities that are given by the (pure) simulation data is excellent for all transitions that are displayed here. Again, we have observed similar agreement for other transitions and different interaction parameters.

In summary, we can conclude that both the implied timescales and the Chapman-Kolmogorov test indicate an excellent quality of the estimated Markov models, suggesting that the assumption of Markovian dynamics on the discretized states should be very well fulfilled. In what follows, we use a lag time of $\tau=5$\,MC~steps to estimate Markov models of the $6^2$ sized Ising system and a lag time of $10\,$MC~steps for larger system sizes.

\subsection{Free energies}\label{sec:MSMfreeEnergies}
\begin{figure}
	\includegraphics[width=\picsize]{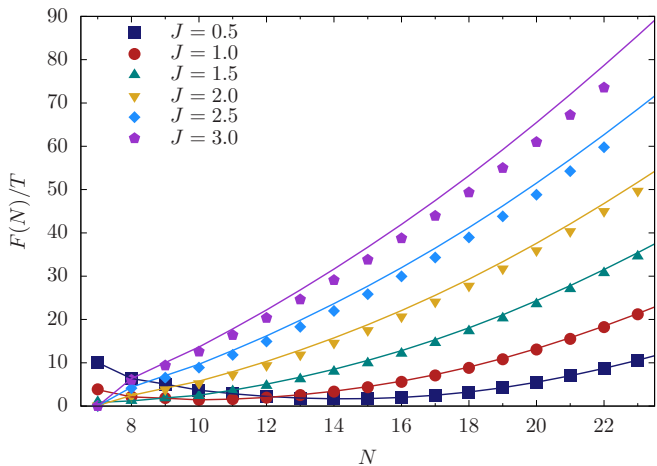}
	\caption{Free energies $F/T$ per order parameter $N$ in systems with varying interaction parameter, computed from the equilibrium probabilities of the corresponding Markov state model \update{(colored points) and computed via WL sampling (colored lines), respectively.}}
	\label{fig:freeEnergies6}
\end{figure}

The free energy of any system is composed of the energetic contribution given by the system's Hamiltonian and an entropic contribution that gains weight with increasing temperature: $F=H-T\,S$. This interplay between energy and entropy exists also in the Ising system, where a variation of the interaction parameter $J$ is equivalent to a variation of the temperature: $F(N)/T=H(N)/T-S(N) = J\, 2 N - S(N)$. Fig. \ref{fig:freeEnergies6} displays the dependence of $F/T$ on the order parameter as derived from the normalized equilibrium probabilities in the estimated Markov state models (see Sect. \ref{sec:methods-SimDetails}) for systems of size $m^2=6^2$. 

The WL algorithm, which we employ as a comparison sampling method, computes the density of states per energy level of a system, which translates to the density of states per order parameter, $g(N)$, in the Ising system. Since the free energy of a certain order parameter is related to the density of states of this order parameter via $F(N)/T = -\ln(g(N))+H(N)/T = -\ln(\pi(N))$, we can compute the free energies from the WL results, too. \update{Fig. \ref{fig:freeEnergies6} also displays the free energies computed via WL sampling, where we have again used normalized equilibrium probabilities to obtain the same relative scale of free energies as in the MSM results.}

The free energy curves for high interaction parameters show a minimum at the minimal order parameter $N=7$, which corresponds to the perfectly phase-separated configuration, and a strongly increasing free energy with increasing order parameter, while the curves for lower $J$ are less steep and the minimum is shifted towards intermediate $N$. This leads to the overall phase-separating behavior in all systems and also demonstrates the interplay between the energetic and entropic contributions to the free energy. While the energetic part dominates at high $J$ and moves the free energy minimum towards low energy configurations with small order parameter, the entropic part gains influence at lower $J$ and the free energy minimum is shifted towards configurations with higher entropy. It is intuitively clear that configurations with medium order parameter $N=\frac{m^2}{2}$ will have the highest entropy simply because an intermediate number of $+-$ pairs can be generated by many more possible arrangements of spins on the lattice than a small or a very high number of those pairs.

\update{The free energy gain upon unmixing can be easily computed as the free energy difference between mixed and phase-separated configurations. As already implied  by Fig. \ref{fig:freeEnergies6}, this free energy gain increases with increasing interaction parameter between the spins and is proportional to the size of the system (see Supplementary Material for additional data with larger system sizes). }

The comparison between the WL results and the MSM estimates for the free energy differences reveals an excellent agreement at low interaction parameters but increasing deviations between both methods at higher interaction parameters. Although these differences seem to be small on an overall scale in Fig. \ref{fig:freeEnergies6}, they will amount to approx. one order of magnitude for $m=6$ in the respective entropies of the systems and increase for larger systems (see below). Given the perfect findings in the tests of Markovianity, this observation may seem very surprising. In what follows, we will give a more detailed comparison between the MSM estimates and the WL results and scrutinize the causes for these deviations at high $J$ values.

\subsection{Detailed comparison with WL results}\label{sec:DetailedCompWL}
Both the density of states and the free energy are directly related via the equilibrium probabilities from MSM (see eq. \ref{eq:freeEnergy} and \ref{eq:Boltzmann}). Therefore, the most straightforward way to directly compare MSM and WL sampling consists of comparing the resulting densities of states (or the respective entropies per $N$, since $g(N)=\exp{S(N)}$). Here, we will again focus on the small $6^2$ sized Ising system but emphasize that we have observed analogous behavior for larger system sizes.

At this point we would like to stress that although a system's density of states is in general temperature-dependent, this is not the case in the Ising model. Here, all microstates within a Markov state have by construction the same energy. Thus, the density of states for a given order parameter must be independent of the interaction parameter between the spins, which only influences the magnitude of the energy differences between states.

\begin{figure}
	\includegraphics[width=\picsize]{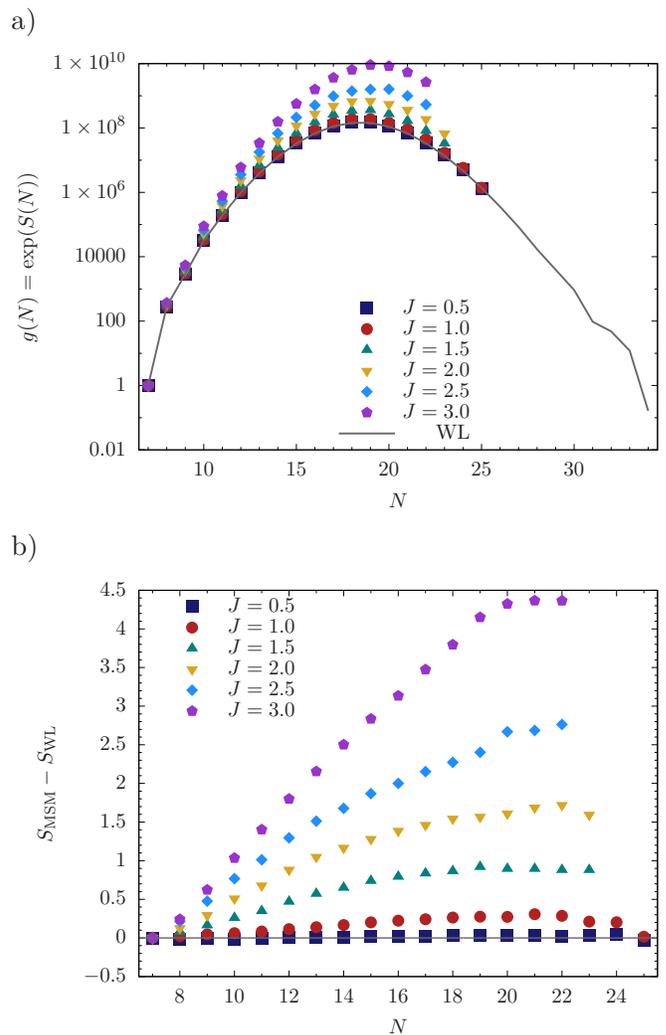}
	\caption{a) Density of states $g(N)$ per order parameter $N$ computed from the equilibrium probabilities in MSM for varying interaction parameter $J$ (colored symbols) and estimated via the WL algorithm (grey line) in systems of size $m^2=6^2$. b) Difference in the entropies computed from the MSM to the entropies from WL sampling. A small N-dependence of this curve indicates good agreement between the WL result and the MSM result, whereas a stronger N-dependence reflects deviations between both methods for the respective range of order parameters.}
	\label{fig:entropies6}
\end{figure}

The upper plot in Fig. \ref{fig:entropies6} displays the densities of states $g(N)$ in a system of size $m^2=6^2$ computed via MSM and WL sampling.
First of all one can notice that the density of states computed with the WL method spans a larger order parameter range than the results from MSM. A Markov state model can only be built between Markov states that have been sampled in the original simulation data. Thus, the high-$N$ states, which correspond to very high free energies and are therefore rarely visited during standard MC simulations, are not included. By contrast, the WL algorithm performs a random walk in energy space and not in configuration space and forces the system to visit all possible energies, i.e. all possible order parameter states in the Ising system, to  estimate the density of states. 

We would like to remark in passing that the kink in the density of states curve in the high order parameter limit and the fact that the curve is not perfectly symmetric, as it would be for a "normal" Ising system\cite{WangLandau2001PRL}, is related to the constraint of equal numbers of up- and down spins on the lattice. 

The density of states from the Markov model for $J=0.5$ and the result from WL sampling agree very well. In contrast, for large $J$ the previous disagreement between the MSM- and the WL-data for the free energy translates into corresponding deviations in the density of states. Due to the imposed shift of all curves to $g(N_\mathrm{min})=1$, these deviations appear to be stronger at higher $N$. The bottom plot in Fig. \ref{fig:entropies6} reveals the true location of these deviations by showing the difference between the entropies from MSM and the entropies from WL sampling as a function of $N$, which is equivalent to the ratio of the corresponding densities of states in a logarithmic representation. The imposed shift only influences the absolute magnitude of this difference and not the overall behavior of the resulting curve. For example, a difference in the entropies of about $3.9$ corresponds to an MSM estimate for the density of states that is fifty times larger than the WL estimate. The slope of this curve, $\frac{\diff}{\diff N} \Delta S$ with $\Delta S = S_\mathrm{MSM}-S_\mathrm{WL}$, indicates the agreement between the MSM results and the WL results. At low interaction parameters, the MSM method gives good estimates for the density of states of the system but the increasing slope in the curves for higher interaction parameters points to significant deviations of the MSM estimates for the densities from the correct density of states in the system. Interestingly, the N-dependence of $\frac{\diff}{\diff N} \Delta S$ is largest for intermediate order parameters (around $N=12$) and becomes smaller close to the maximum of the entropy curve ($16 \le N \le 22$) and in the limit of low order parameters ($N = 7,8$).

In summary, these results clearly indicate the presence of subtle non-Markovian effects, giving rise to systematic errors in the MSM estimation process, despite a positive validation process for the Markov models. It suggests that the partitioning scheme that we used, which groups \update{configurations according to their degree of phase separation, i.e. according to their number of $+-$-pairs}, is not sufficiently fine in order to resolve the important dynamical processes of the system on the discretized level. Hence, a couple of questions emerge. What distinguishes the configurations in the Ising system apart from \update{their number of $+-$-pairs, i.e. their energies,} and why did the validation methods fail to reveal the existing problems with the Markov models? Can we reasonably enhance the Markov models by improving the partitioning scheme or do these erroneous MSM estimates at higher $J$ have a more fundamental origin?

\section{Refining of the discretization scheme}\label{sec:SubdivisionNstates}
 Here we proceed by introducing a further partitioning of the previous order parameters. Although no relevant improvement in the MSM estimate is observed, we gain new insight into possible reasons for the deviations, which will be explored in the subsequent Section.

\subsection{Subdivision of the order parameter states}
\begin{figure}
	\includegraphics[width=\picsize]{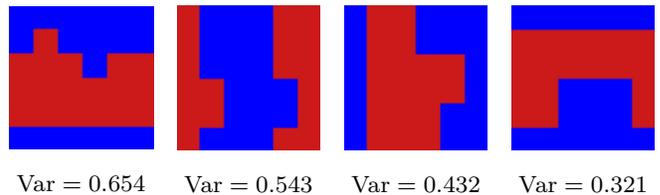}	
	\caption{Example configurations that depict the fact that not all configurations with the same order parameter have the same variance $\mathrm{Var}(M_{+-})$. Here, all four possible variance types that correspond to $N=8$ in a $6^2$ sized Ising system are illustrated by example configurations.}
	\label{fig:varianceStates}
\end{figure}

Until now, we have grouped together all configurations that have the same number of $+-$ pairs. A natural refinement of the order parameter would be to additionally take into account the variance of the number of $+-$ pairs on the lattice, i.e.
\begin{align}
	\mathrm{Var}(M_{+-}) &= \langle M_{+-}^2 \rangle - \langle M_{+-} \rangle ^2 \\ \notag
	&= \frac{1}{n} \left( \sum_i M_{+-}^2(i) - \left(\sum_i M_{+-}(i)\right)^2 \right).
\end{align}
Here, $M_{+-}(i)$ denotes the number of $+-$ pairs for spin $i$ and $n$ is the total number of spins. Not all configurations of a given order parameter have the same $+-$ variance, as can be seen in Fig. \ref{fig:varianceStates}. There, four configurations of a $6^2$ sized system are shown which all have order parameter $N=8$ but different values of the variance. However, the possible values of this variance are also very restricted due to the lattice structure of the system. We can therefore distinguish between groups of configurations that correspond to the same order parameter according to their variance.

\subsection{Non-equilibrium dynamics}
\begin{figure}
	\includegraphics[width=\picsize]{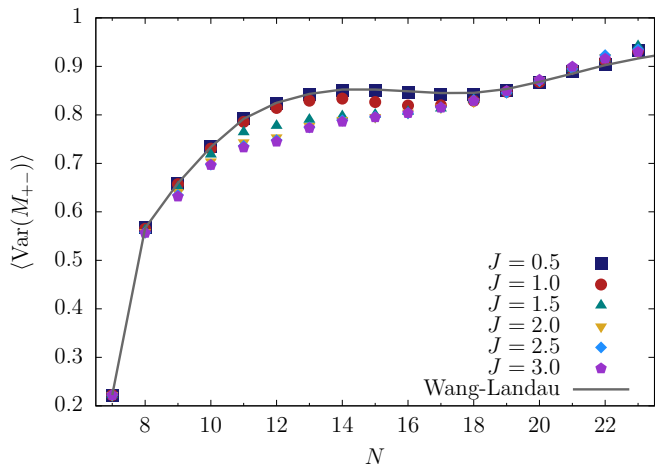}
	\caption{Average $+-$ variance per order parameter computed during short, non-equilibrated classical MC simulations with varying interaction parameter and during a WL simulation.}
	\label{fig:variancePlot}
\end{figure}

Using the (pure) MC simulation data, we also analyzed the average $+-$ variance as a function of the order parameter; see fig. \ref{fig:variancePlot}. Interestingly, one finds again a significant $J$-dependence. The strongest $J$-dependence is observed for intermediate $N$-values, for which also the $J$-dependence in MSM was strongest (see Sect. \ref{sec:DetailedCompWL}). For small $J$ the data agrees with the Wang-Landau results. In contrast, for large $J$ the average $+-$ variance is smaller than expected from equilibrium. Of course, this dependence has to disappear for global equilibrium. Indeed, when increasing the length of the simulations, the $J$-dependence becomes smaller, in particular for small order parameters, i.e. around the minimum of the free energy curve. However, for any realistic simulation times one would not observe that unmixed systems are mixed again, thereby generating thermodynamic equilibrium. This result shows, first, that the $J$-dependence in the MSM analysis is not an artifact of our procedure and, second, that a closer understanding of the relevance of the $+-$ variance may shed light on the underlying reason of our surprising observations.

If the underlying non-Markovian properties were exclusively related to the presence of states with different $+-$ variances, a refined partitioning of the configuration space according to order parameter \emph{and} $+-$ variance should be void of any $J$-dependence. As shown in the Supplementary Material, this $J$-dependence is slightly reduced but is still significant when using this refined partitioning scheme in MSM. \update{Hence, the non-Markovianity effects are not fully compensated by using this refined discretization scheme in MSM but the fact that they are significantly reduced suggests that the partitioning refinement incorporates some of the important distinctions between the configurations.}

\subsection{Connectivity analysis}
\begin{figure}
	\includegraphics[width=\picsize]{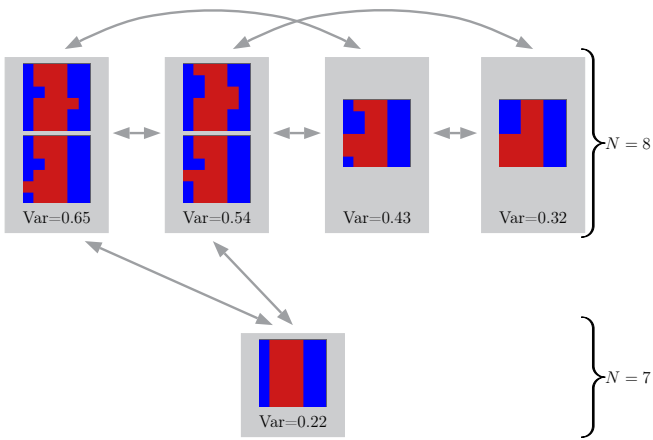}
	\caption{Depiction of the connectivity between the configurations with lowest energies in a $6^2$ Ising system. Configurations are grouped according to their order parameter and their  $\mathrm{Var}(M_{+-})$ (grey boxes) and exemplary configurations of each group are shown. A grey arrow between groups signifies that there is at least one configuration in the first group that can be converted to a configuration within the second group by a single MC move.}
	\label{fig:connectivityPlot}
\end{figure}

Now, we provide evidence that an important distinction between these (energetically equal) configurations with different $+-$ variance is the connectivity between them. This connectivity is related to the chosen move class of the MC simulation. Because of the huge total number of configurations that exist already in the small $6^2$ system, we settle for a closer look at the connectivity between configurations with low energy, namely those with order parameter $N=7,8$, assuming that the behavior at higher energies is admittedly more complex but somewhat analogue. We make use of the partitioning of configurations according to the order parameter and the variance $\mathrm{Var}(M_{+-})$ in order to illustrate the resulting connectivity in Fig. \ref{fig:connectivityPlot}, keeping in mind that this is still a coarse grained picture.
The analysis of the connectivity does indeed reveal that not all isoenergetic configurations are equivalent when it comes to connectivity: on the one hand, not all different $\mathrm{Var}(M_{+-})$ configurations within one order parameter are directly connected to each other, e.g. the configurations with $\mathrm{Var}(M_{+-})=0.65$ cannot convert to a configuration with $\mathrm{Var}(M_{+-})=0.32$ with just one MC step. On the other hand, not all configurations with $N=8$ are directly connected to the $N=7$ configuration, this is only given for the $N=8$ configurations with $\mathrm{Var}(M_{+-})=0.65$ or $0.54$, respectively. Moreover, this second statement is also only true for specific configurations within the  $\mathrm{Var}(M_{+-})=0.54$ group. For example, the upper configuration of the $N=8$ and $\mathrm{Var}(M_{+-})=0.54$ group in Fig. \ref{fig:connectivityPlot} cannot be converted to a $N=7$ configuration with only one MC step, which shows that the true connectivity between all configurations is even more complex.

\section{Analytic solution of a five state model system with absent connections}\label{sec:5statemodel}
\begin{figure}
	\includegraphics[]{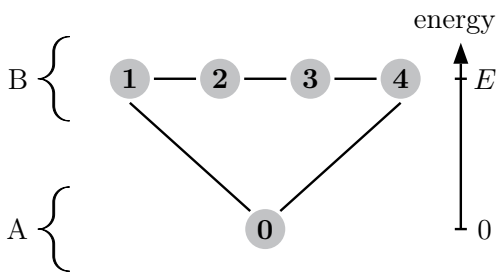}
	\caption{Depiction of a model system containing five distinguishable microstates which are grouped into two Markov states, termed A and B, according to their energy. The lines represent the connectivity between the microstates.}
	\label{fig:5stateModel}
\end{figure}

Here, we show that absent connections between a microstate of one Markov state to any microstates in another Markov states may indeed cause erroneous MSM estimates and that the deviations will depend on $J$. For the illustration of this mechanism we consider a minimalistic five state model (see Fig. \ref{fig:5stateModel}), which can be analyzed in great detail.

\subsection{Derivation of the analytic solution}
The model contains five microstates of which four have energy $E$, with $E>0$, and are numbered 1 to 4 and the fifth microstate, termed microstate 0, has an energy of zero. These microstates are grouped into two Markov states according to their energy, so that Markov state B contains all four microstates with energy $E$ and Markov state A is formed by the microstate with energy zero. The probability to accept a transition from microstate 0 to microstate 1 or 4, respectively, is given by $\varepsilon = e^{-\beta E}$ in the Metropolis Monte Carlo scheme, whereas all other possible transitions are always accepted because they either lower the energy of the system or keep it unchanged. The important part of this model consists of the fact that two of the microstates in B are not directly connected to Markov state A: if the system is in microstate 2 (or 3) it has to first perform a step within B and jump to microstate 1 (or 4) before it can perform a transition to Markov state A. This feature is supposed to describe the effect of absent connections between microstates that we have observed in the Ising system.

The ratio between the equilibrium probabilities of both Markov states, $\pi_\mathrm{B}/\pi_\mathrm{A}$, is given by their Boltzmann probabilities
\begin{align}
	\pi_\mathrm{B}=\frac{4\varepsilon}{Z}, ~ \pi_\mathrm{A} = \frac{1}{Z} ~~~\mathrm{s.t.}~~~ \frac{\pi_\mathrm{B}}{\pi_\mathrm{A}}=4\varepsilon, \label{eq:probabilities}
\end{align}
where $Z$ denotes the partition function of the system.

Assuming that the probabilities for the microstates are denoted $x_0, x_1, ..., x_4$, the symmetry of the model system allows to condense these five variables to three variables according to $y_0=x_0$, $y_1 = x_1 + x_4$ and $y_2 = x_2 + x_3$. Then, the probability vector, $\mathbf{y}$, at time $t+1$ is given by the matrix product of the transition probability matrix and the probability vector at time $t$:
\begin{align}
	\mathbf{y}^T(t+1) = \mathbf{y}^T(t) \, \mathbf{T} = \begin{pmatrix} y_0 & y_1 & y_2  \end{pmatrix} \begin{pmatrix} 1-\varepsilon &  \varepsilon & 0 \\ \frac{1}{2} & 0 & \frac{1}{2} \\ 0 & \frac{1}{2} & \frac{1}{2} \end{pmatrix}   \label{eq:pmalMatrix}
\end{align}
Here, $T_{ij}$ denotes the probability to transition from state $i$ to state $j$ given that the system is in state $i$ and can be derived from the structure of the model system. For example, the probability $T_{01}$ is equal to $\varepsilon$ because the transition attempt will be accepted with probability $\varepsilon$ and $T_{12} =1/2$ reflects the fact that half of the transition attempts out of state 1 are going to state 2 which, furthermore, are always accepted. For simplicity, we use a lag time of one MC step.
The evolution of the system after $k$ MC steps is then given by
\begin{align}
	\mathbf{y}^T(t+k) = \mathbf{y}^T(t)\,\mathbf{T}^k\label{eq:evolyvec}
\end{align}
and the stationary distribution of the system corresponds to the left eigenvector of the transition matrix $\mathbf{T}$ to the eigenvalue 1, $\mathbf{ev}_1=(1/Z, 2\varepsilon/Z,2\varepsilon/Z)$.

In the Markov modeling scheme, only transitions between Markov states are distinguished. Here, the trivial estimator of the probability for a transition from A to B is given by the number of observed transitions, $C_\mathrm{AB}$, divided by the total number of transitions out of state A, $C_\mathrm{A}$:
\begin{align}
	p_\mathrm{AB} = \frac{C_\mathrm{AB}}{C_\mathrm{A}}~, \mathrm{~~~with~~~} C_\mathrm{A}= C_\mathrm{AA} + C_\mathrm{AB}.
\end{align}
Similarly, the probability for a transition from B to A is given by
\begin{align}
	p_\mathrm{BA} = \frac{C_\mathrm{BA}}{C_\mathrm{B}}~, \mathrm{~~~with~~~} C_\mathrm{B}= C_\mathrm{BB} + C_\mathrm{BA}.
\end{align}
Using these relations together with the condition for detailed balance, $\pi_\mathrm{A}p_\mathrm{AB}=\pi_\mathrm{B}p_\mathrm{BA}$, results in a ratio of the stationary probabilities of
\begin{align}
	\frac{\pi_\mathrm{B}}{\pi_\mathrm{A}} = \frac{C_\mathrm{AB}\,C_\mathrm{B}}{C_\mathrm{BA}\,C_\mathrm{A}}. \label{eq:piBpiA}
\end{align}
We can simplify this relation if we use some information about the structure of the system, i.e. if we exploit the fact that Markov state A consists of only one microstate and that a fraction $\varepsilon$ of the attempted transitions are accepted, i.e. $C_\mathrm{AB}=\varepsilon\,C_\mathrm{A}$. This leads to
\begin{align}
	\frac{\pi_\mathrm{B}}{\pi_\mathrm{A}} = \varepsilon \left( 1 + \frac{C_\mathrm{BB}}{C_\mathrm{BA}}\right).
\end{align}
The number of transitions during one MC step can be related to the probabilities of the microstates at the beginning of that step:
\begin{align}
	\overset{\sim}{C}_{\rm BB}(t+1) &=  y_1(t)+2y_2(t) \\
	\overset{\sim}{C}_{\rm BA}(t+1) &=  y_1(t)
\end{align}
and the resulting numbers after multiple MC steps are then given by summing over all MC steps:
\begin{align}
	C_\mathrm{BB}(t) &= \sum\limits_{j=1}^{t} \overset{\sim}{C}_\mathrm{BB}(j) = \sum\limits_{i=0}^{t-1} \left( \frac{1}{2} y_1(i)+ y_2(i) \right) \label{eq:Csum1}\\
	C_\mathrm{BA}(t) &= \sum\limits_{j=1}^{t} \overset{\sim}{C}_\mathrm{BA}(j) = \sum\limits_{i=0}^{t-1} \frac{1}{2} y_1(i). \label{eq:Csum2}
\end{align}
Thus, the MSM estimate of $\pi_\mathrm{B}/\pi_\mathrm{A}$  can be computed by calculating the evolution of an initial probability vector, $\mathbf{y}(0)$, at a series of MC steps with eq. \eqref{eq:evolyvec} and then using these probabilities to compute the $C_\mathrm{BB}(t)$ and $C_\mathrm{BA}(t)$ with eqs. \eqref{eq:Csum1} and \eqref{eq:Csum2}.

In order to compute the implied timescale of the five state model, this scheme has to be extended to lag times $\tau > 1$ (see Appendix \ref{app:Derivation_tau}). Then, the implied timescale $t_2$ which would be observed in MSM can be computed via $t_2=-\tau/\ln(\lambda_2)$ from the non-unity eigenvalue, $\lambda_2$, of the normalized row-stochastic count matrix 
\begin{align}
	\mathbf{C}(t,\tau) = \left( \begin{matrix}
							\frac{C_\mathrm{AA}(t,\tau)}{C_\mathrm{A}(t,\tau)}	& \frac{C_\mathrm{AB}(t,\tau)}{C_\mathrm{A}(t,\tau)} \\
							\frac{C_\mathrm{BA}(t,\tau)}{C_\mathrm{B}(t,\tau)}	& \frac{C_\mathrm{BB}(t,\tau)}{C_\mathrm{B}(t,\tau)}
						 \end{matrix} \right),
\end{align}
where $t$ denotes the number of MC steps that are used to compute the transition counts.

\subsection{Results}\label{sec:5statemodel-results}
\begin{figure}\centering
	\includegraphics[]{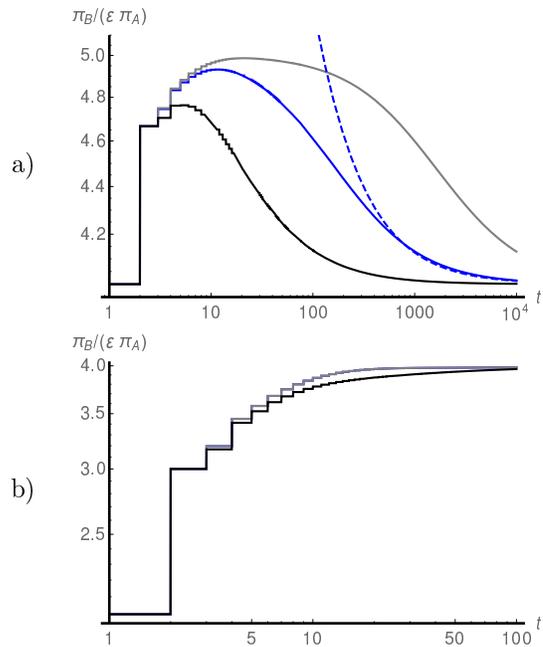}
	\caption{Evolution of the ratio $\frac{\pi_\mathrm{B}}{\varepsilon \pi_\mathrm{A}}$ with the number of MC steps for an initial probability vector of $\mathbf{y}^T(0)=(0,\frac{1}{2}, \frac{1}{2})$ (a) and $\mathbf{y}^T(0)=(0,1,0)$ (b), respectively. The color coding corresponds to different values of $\varepsilon$ according to: blue $\rightarrow \varepsilon =\exp{-5}$, grey $\rightarrow \varepsilon / 10$, black $\rightarrow \varepsilon \times 10$ and the dashed curve denotes a fitting function that is explained in the text.}
	\label{fig:5stateAnalyticSols}
\end{figure}

Fig. \ref{fig:5stateAnalyticSols} displays the evolution of the estimator for $\pi_\mathrm{B}/(\varepsilon \pi_\mathrm{A})$ as a function of the number of MC steps, which will be denoted as $\hat{\pi}(t)$ hereafter, for different initial populations, choosing $\varepsilon=\exp{-5}$ (corresponding to $J=2.5$ in the Ising system). In the Fig. at the top (case I), the time evolution starts from a local equilibrium distribution in Markov state B, at the bottom (case II) from a population which is initially located only in the "edge" microstates 1 and 4. We have also added simulations where $\varepsilon$ was either increased or decreased by a factor of 10. In equilibrium one expects $\hat{\pi} = 4$ (see eq. \eqref{eq:probabilities}). Indeed, this limit is reached in both cases for long times. Since the initial population in case I represents the local equilibrium, the estimation after one step has to correspond to the equilibrium one \cite{Noe2011}, in agreement with Fig. \ref{fig:5stateAnalyticSols}. Afterwards, the estimate departs from the equilibrium value because the population of the upper levels is no longer in local equilibrium after the first time step. Because the initial population in case II is a non-equilibrium distribution, the estimator starts with a wrong value but converges towards the equilibrium value at longer times, as expected.

\begin{figure}\centering
	\includegraphics[]{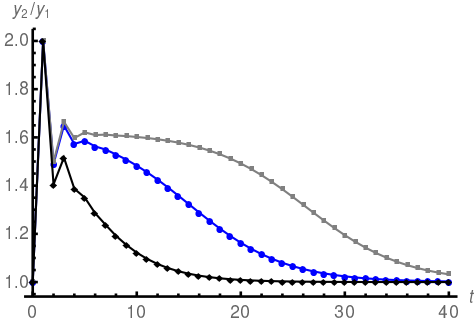}
	\caption{Evolution of the ratio $y_2(t)/y_1(t)$ for case I (initial population $\mathbf{y}^T(0)=(0,\frac{1}{2}, \frac{1}{2})$) with the number of MC steps. The color coding corresponds to different values of $\varepsilon$ according to: blue $\rightarrow \varepsilon =\exp{-5}$, grey $\rightarrow \varepsilon / 10$ and black $\rightarrow \varepsilon \times 10$.}
	\label{fig:5stateAnalyticSols-y2y1}
\end{figure}

Interestingly, we find three very different timescales which determine the time evolution of our system. First, the global equilibration time is determined by the slowest relaxation process of the system, i.e. by its largest implied timescale. In the limit of high energy differences or low temperatures, respectively, corresponding to $\varepsilon \ll 1$, this timescale is about $t_\mathrm{equil}\approx 5$~MC~steps and does not depend on $\varepsilon$; see Appendix \ref{app:Derivation_tequil}. Thus, one might intuitively expect that equilibrium is restored in all dynamic variables for $t > t_\mathrm{equil}$. In particular, one might expect that (i) the ratio $y_2(t)/y_1(t)$ approaches one and (ii) the estimator $\hat{\pi}(t)$  approaches four. However, both propositions are generally not correct. Specifically, we discuss the initial condition of case I. In Fig. \ref{fig:5stateAnalyticSols-y2y1} we show the corresponding ratio of $y_2(t)/y_1(t)$ for different values of $\varepsilon$. A straightforward analytic calculation, discussed in Appendix \ref{app:Derivation_tequil}, shows that this ratio is approx. 1.6 for times smaller $t_\mathrm{equil,1} \approx t_\mathrm{equil} \times \ln(\frac{1}{4\varepsilon})$ and then approaches its equilibrium value of one. Basically, transitions from Markov state A to Markov state B become important around $t \approx t_\mathrm{equil,1}$ to establish the equilibrium ratio $y_2(t)/y_1(t) = 1$. An even longer time $t_\mathrm{equil,2}$ is required (see upper plot in Fig. \ref{fig:5stateAnalyticSols}), so that the estimator $\hat{\pi}$ converges to its equilibrium value. Thus, the reinstallment of local equilibrium is \emph{not} sufficient to correctly estimate the Boltzmann distribution. Indeed, this effect becomes even larger for higher energy differences, as also shown in Fig. \ref{fig:5stateAnalyticSols}. The numerical data suggests a scaling of $t_\mathrm{equil,2}$ by $1/\varepsilon$. Numerically, we observe that $\hat{\pi}$ yields a maximum of exactly five (instead of four) for very small $\varepsilon$ as long as $t \ll t_\mathrm{equil,2}$. In general, this long convergence time occurs if $y_2(0) \ne 0$. Hence, the estimator $\hat{\pi}$ approaches its equilibrium value on the timescale of $t_\mathrm{equil}$ in case II. Please note that for very small $\varepsilon$, relevant at low temperatures, one has $t_\mathrm{equil} \ll t_\mathrm{equil,1} \ll t_\mathrm{equil,2}$.

The relation $t_\mathrm{equil} \ll t_\mathrm{equil,2}$ can be easily understood. First, we note that the ratio of the number of transitions per MC step, $\overset{\sim}{C}_{\rm BB}$ and $\overset{\sim}{C}_{\rm BA}$, converges to its equilibrium value of 3 on the timescale of $t_\mathrm{equil,1}$ in analogy to $y_2(t)/y_1(t)$ (see Supplementary Material). Analyzing both values independently, we observe that they are of the order of one for short times because the system starts in the upper Markov state. For times larger than $t_\mathrm{equil,1}$, they are of the order of $\varepsilon$. The number of observed transitions $C_\mathrm{AB}$ and $C_\mathrm{BB}$ are basically the sums over the $\overset{\sim}{C}_{\rm AB}$ and $\overset{\sim}{C}_{\rm BB}$, respectively. Thus, if the estimator $\hat{\pi}$ deviates from the equilibrium value on a time, say $t=t_\mathrm{equil,1}$, the subsequent convergence to the correct value requires an additional time interval of the order of $1/\varepsilon$ because the additional contributions in the sums are so small.

These arguments can also be rephrased in a more quantitative way.
Under the assumption that all microstates have acquired their equilibrium values for $t > t_\mathrm{equil,1}$, one has $\overset{\sim}{C}_{\rm BB} \approx 6 \epsilon$ and $\overset{\sim}{C}_{\rm AB} \approx 2 \epsilon$. In this time regime, this yields $C_\mathrm{BB}(t) \approx C_\mathrm{BB}( t_\mathrm{equil,1}) + 6 \varepsilon t$ and $C_\mathrm{BA}(t) \approx C_\mathrm{BA}( t_\mathrm{equil,1}) + 2 \varepsilon t$, giving rise to the approximation
\begin{align}
\hat{\pi}=\pi_\mathrm{B}/(\varepsilon \pi_\mathrm{A}) &= 	1+\frac{C_\mathrm{BB}(t)}{C_\mathrm{BA}(t)} \nonumber \\ & \approx 4+(C_\mathrm{BB}(t_\mathrm{equil,1})-3C_\mathrm{BA}(t_\mathrm{equil,1}))\frac{1}{2\varepsilon t}.
\end{align}
for $t \ge 1/\varepsilon$.
As can be seen in Fig. \ref{fig:5stateAnalyticSols} (denoted as "fit" and computed with $t_\mathrm{equil,1} \approx 20 $\,MC steps), this describes the numerical data very well for long times.

This long convergence process becomes relevant if the ratio $C_\mathrm{BB}/C_\mathrm{AB}$ still differs from its equilibrium value on the time scale of a few relaxation times $t_\mathrm{equil}$. As indicated by this simple model, the effect becomes relevant if microstates are initially populated which are not connected to microstates with lower energy and therefore acquire too much weight during the time of their population.

More generally, one may consider a system with two Markov states with $n_\mathrm{A}$ and $n_\mathrm{B}$ microstates, each microstate having connections to $k$ randomly chosen microstates. Then, one expects that the described estimation bias becomes relevant if there is a sufficient number of microstates in Markov state B which are not connected to any microstate in Markov state A. This statement is equivalent to $k \ll n_\mathrm{B}/n_\mathrm{A}$. As a consequence, the bias will become smaller if the ratio $n_\mathrm{B}/n_\mathrm{A}$ becomes smaller. \update{Furthermore, one expects a dependence on the parameter $n_\mathrm{B}/(k\,n_\mathrm{A})$. The corresponding results are shown in the Supplementary Material.} 

\begin{figure}\centering
	\includegraphics[]{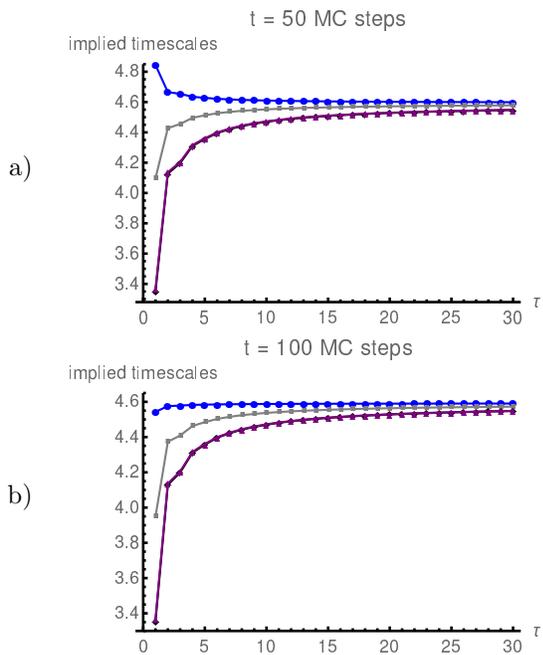}
	\caption{Implied timescales of the five state model system for different total numbers of MC steps (50 MC steps (a) and 100 MC steps (b), respectively) and various initial population vectors: $\mathbf{y}^T(0)=(0,0,1)$ (blue), $\mathbf{y}^T(0)=(0,\frac{1}{2},\frac{1}{2})$ (grey), $\mathbf{y}^T(0)=(0,1,0)$ (black) and the globally equilibrated population $\mathbf{y}^T(0)=(\frac{1}{Z},\frac{2\varepsilon}{Z},\frac{2\varepsilon}{Z})$ (purple). Please note that the black curve is mostly overlaid by the purple curve.}
	\label{fig:5stateAnalyticSols-ITS}
\end{figure}

Finally, the question whether this estimation bias would be noticeable in the standard MSM validation methods, such as, e.g., the behavior of the implied timescales with varying lag times, remains to be studied. For that purpose, Fig. \ref{fig:5stateAnalyticSols-ITS} displays the implied timescales as a function of $\tau$ in the five state model for various initial populations and trajectory lengths of $60$ and $100$ MC steps, respectively. A constant behavior of the implied timescales with varying lag time is usually seen as an indicator for Markovianity in the corresponding Markov state model\cite{Swope2004, Chodera2006}. 
Very surprisingly, the implied timescales show approximately constant behavior with varying lag times when starting with an initial distribution that has significant population in those microstates that are not directly connected to Markov state A ($\mathbf{y}^T(0)=(0,0,1)$) and an increasing dependence on the lag time with decreasing initial population in those microstates. In those cases with significant initial population in the "edge" microstates ($\mathbf{y}^T(0)=(0,1/2,1/2)$ and $\mathbf{y}^T(0)=(0,1,0)$), approximately constant implied timescales are reached only at $\tau > 5$ MC steps. Consequently, the estimation bias becomes marginal when estimating the Markov state model at larger lag times (see Supplementary Material). Thus, taking the standard procedure one would have guessed that Markovian behavior is present for very small lag times. This is in marked contrast to the actual data from Fig. \ref{fig:5stateAnalyticSols} and \ref{fig:5stateAnalyticSols-y2y1}. In other words and contradicting to the existing beliefs, one could say that the larger the estimation bias, the more would the corresponding implied timescales plot suggest perfectly Markovian behavior. 

\subsection{Relevance for the MSM analysis of unmixing}
In the Ising model and in particular at large $J$, up-transitions most likely occur only for adjacent energy states. Therefore, we may conceptually regard the Ising model as a succession of pairs of neighboring Markov states which individually follow the general behavior of the model system, at least on a qualitative level. Here, we discuss the implications of this relation for the results of the Ising model.

As explicitly shown above for the connectivity between states with $N=7$ or $N=8$, there exist groups of configurations (e.g. those with +- variance 0.32)  with no direct connection to configurations with lower $N$. If we may generalize this observation to all pairs of neighboring order parameters, the properties of the five state system are likely relevant for all pairs of Markov states in the Ising model. Starting from some (high) energy state, the count matrix will (on average) suggest that the entropic weight of this Markov state is higher than the weight of the energetically lower Markov state. In contrast to the five state model, the system will typically descend in energy (i.e. convert to configurations with lower energy) much before the convergence time $t_\mathrm{equil,2}$ of this pair of Markov states is reached. As a consequence, the slope of the difference function $\Delta S(N) = S(N) - S(N)_\mathrm{true}$, i.e. $\frac{\diff}{\diff N}\Delta S(N)$, resulting from the MSM analysis, is positive. Going from the unmixed state until the free energy minimum, this effect accumulates and gives rise to the systematic underestimation of the free energy difference that drives the phase separation.

The results of the five state model also suggest that this artificial increase should be less prominent in three different cases:
(i) around the low-energy regime, $\frac{\diff}{\diff N}\Delta S(N)$ becomes much smaller because the system can fully equilibrate close to the minimum of the free energy. Thus, in analogy to the insight from the five state model, one has sufficient time to reach convergence of the number of transitions. In this case the error is strongly reduced.
(ii) Furthermore, the  five state model can explain why the entropy predictions are much better for smaller values of $J$. First, the pair-wise convergence is much faster for smaller energy differences. Second, the thermodynamic driving force of unmixing is much weaker so that the system may stay somewhat longer in a given energy region. Third, since the probability that a MC move increases the order parameter by two or more increases with decreasing $J$, an even stronger connectivity is reached.
(iii) Let $f(N)$ denote the ratio of the absolute number of states in two successive $N$ levels, i.e. $f(N) = \exp{S(N) - S(N-1)}$. For our model analysis we have seen that the estimation bias is stronger for larger $f(N)$, corresponding to a surplus of high-energy states. Close to the region of maximum entropy, corresponding to statistically mixed configurations, $f(N)$ approaches zero. Indeed, we find in the bottom plot of Fig. \ref{fig:entropies6} that $\frac{\diff}{\diff N}\Delta S(N)$ decreases close to the region of maximum entropy.

Finally, we can rationalize why the average $+-$ variance appeared to be smaller for larger $J$, in particular for intermediate order parameters (see Fig. \ref{fig:variancePlot}). In the five state model, we found that microstates in a Markov state with weaker connectivity to the lower Markov state display stronger populations for $t < t_\mathrm{equil,1}$. The analysis of the Ising model (see Fig. \ref{fig:connectivityPlot}) suggests that microstates with smaller variances are less good connected than microstates with higher variance. Both observations together support our numerical findings.

All observations in Sect. \ref{sec:MSMfreeEnergies} and \ref{sec:DetailedCompWL} reflect the non-Markovian nature of transitions between the chosen Markov states of the Ising model. However, the Markov-tests for the Ising model clearly failed to detect these non-Markovian effects. Here again, the five state model helps to understand this effect. As could be seen in Fig. \ref{fig:5stateAnalyticSols-ITS}, trajectories with large estimation bias lead to a more constant behavior of the largest implied timescale with respect to the lag time, suggesting Markovianity, than trajectories with smaller or even no estimation bias, such as a globally equilibrated trajectory. In analogy to the previous considerations, it seems reasonable to expect a similar constancy for the overall implied timescales in the Ising model if every pair of neighboring Markov states displays no dependence on the lag time due to a prevailing estimation bias.

\section{Accounting for absent connections in the MSM estimation process}\label{sec:AbsentConnectionsAlgorithm}
The previous part demonstrated on the basis of a five state model system that absent connections between microstates have a significant impact on the bias in Markov state modeling that is caused by assuming instant local equilibration within the Markov states. Here, we develop an algorithm that is supposed to diminish this bias in MSM of the Ising system by sorting out the transition counts that are extracted from the discretized trajectories.

\subsection{Derivation of the absent-connections-algorithm}
\begin{figure}
	\includegraphics[width=\picsize]{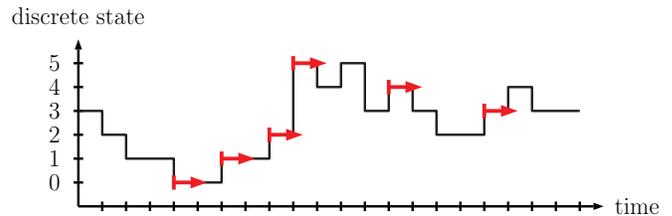}
	\caption{Depiction of a toy discretized trajectory with states $0,1,...5$ that have energies $E_0 < E_1 < ... < E_5$ and the impact of the "absent connections" algorithm: transitions $i\rightarrow j$ are only recorded if the system has previously visited a state $k$ with lower energy. The points in the trajectory from which this criterion has been met for a certain discrete state $i$ are indicated by red arrows. Transitions out of the state with the lowest possible energy are always recorded.}
	\label{fig:missingUpJumpsAlg}
\end{figure}

In this algorithm we exploit the observation that every jump from a lower energy order parameter state to a higher order parameter state will result in a configuration with maximal connectivity, meaning that it is directly connected to both other configurations with the same energy as well as to configurations with lower energy. The previous part showed that trajectories which begin to explore the corresponding phase space region starting from such a configuration with maximal connectivity need significant smaller timescales to recover unbiased MSM estimates of the stationary distribution (case II in Sect. \ref{sec:5statemodel-results}). Thus, we restrict the transitions within one discretized trajectory that are used for the MSM estimation process to those transitions that are preceded by such an "increasing-energy" transition. More specifically, transitions $i \rightarrow j$ are only used for the estimation if there has been a transition $k \rightarrow i$ from a state $k$ with $E_k < E_i$ earlier in the trajectory. We realize that this is a rather weak criterion because it does not discard subsequent transitions from a high energy state $l$ to a low energy state $m$ if there has been a transition from a lower energy state $k$ to $l$ anywhere earlier in the same trajectory. However, we assume that if transitions between states are frequent in a trajectory, as is, e.g., the case between the low energy (unmixed) order parameter states in the Ising system, relaxation towards equilibrium will be fast and the action of the algorithm is not required, in contrast to those transitions between states that are rare in the trajectory and where overestimation of the probabilities due to absent connections will have a large impact. Fig. \ref{fig:missingUpJumpsAlg} illustrates the working principle of this algorithm on the basis of a toy discretized trajectory with Markov states 0,1,...,5 that have increasing energies $E_0 < E_1 < ... < E_5$. \update{Please note that the new algorithm is only marginally more time-consuming that the standard MSM algorithm.}

\subsection{Results and discussion}
\begin{figure}
	\includegraphics[width=\picsize]{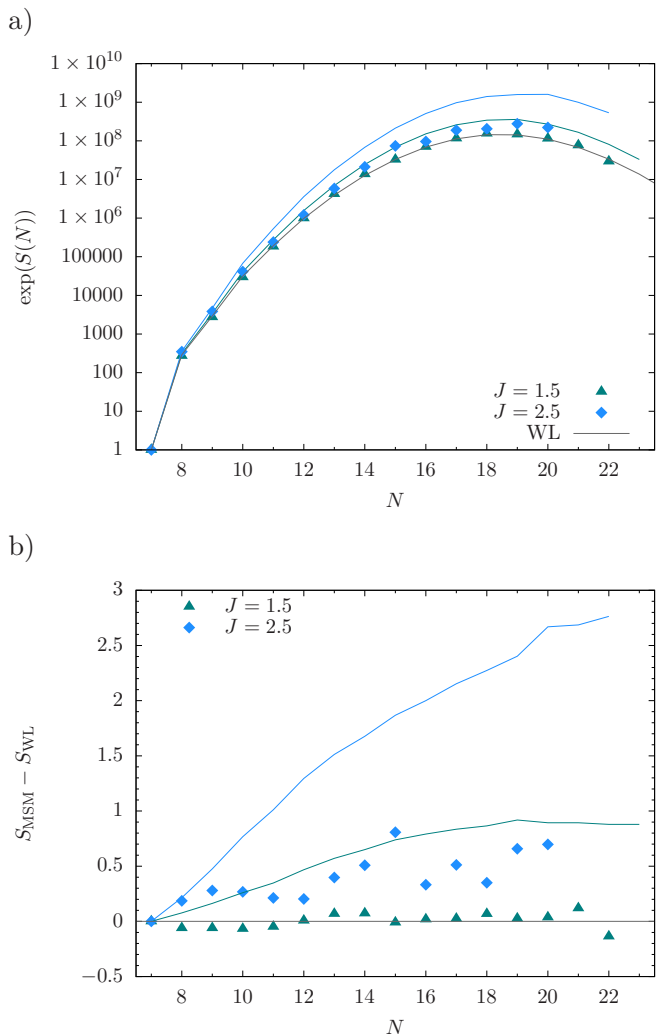}
	\caption{Density of states $g(N)=\exp{S(N)}$ (a) and the difference in the entropies $S_\mathrm{MSM} - S_\mathrm{WL}$ (b) as a function of the order parameter $N$. The colored lines correspond to the results from MSM that have been obtained without using the algorithm, the grey line represents the density of states from the WL algorithm and the colored symbols correspond to the new results from MSM with the new algorithm.}
	\label{fig:entropies6_alg}
\end{figure}

Fig. \ref{fig:entropies6_alg} shows the estimates for the density of states $g(N)$ that result from MSM with this new algorithm as well as the previously presented estimates that were obtained without using the algorithm for a $6^2$ sized system. The comparison shows a significant improvement of the MSM results when this new algorithm is used, especially at moderate interaction parameter values. The curve for $J=2.5$ indicates that the algorithm might still be improvable since this curve shows some regions where the MSM results still deviate from the WL results, but the overall effect of this algorithm to the MSM results is remarkable. Plus, it is worth noting that the application of this algorithm even results in better MSM estimates for the density of states than using the finer $N+\mathrm{Var}$ discretization scheme does (see Supplementary Material).

The improvement of the MSM results is somewhat diminished in larger systems but is still observable \update{(see Supplementary Material for the case of a $10^2$ sized system)}. This observation can be related to the fact that large systems can be conceptually decomposed into weakly coupled subsystems (at least far away from critical points). Imagine that we simulate a large number of independent smaller systems but only observe the sum of all local order parameters. Then, the algorithm would be very inefficient because it would just react to the transitions in one subsystem whereas the transitions in all other subsystems are not influenced and produce somewhat biased data for the transition matrix.

\section{Conclusion} \label{sec:conclusion}
In the present study we have, for the first time, shown that the MSM methodology is fully appropriate to estimate the free energy gain of unmixing based on many short simulations. A closer analysis has revealed some interesting deviations from the theoretical expectation in the limit of high $J$ or, equivalently, low temperatures. They could be traced back to effects of local non-equilibrium, which turn out to be relevant for the present type of problem where the high-energy regions are crossed quite fast. More specifically, the emergence of non-equilibrium populations could be related to the different connectivities of the individual microstates of a Markov state to a Markov state with lower energy. The five state system, analyzed in detail, helps to rationalize the different numerical observations.

Typically, incorrect contributions to the transition matrix occur if microstates are initially populated which are not connected to microstates with lower energy. We have devised an algorithm which is able to strongly reduce this counting problem. The mere fact that this algorithm works so well for the small system is another strong hint that we have identified the correct underlying problem that is giving a systematic bias to the MSM estimates.

\update{In general, one would usually choose a kinetically relevant clustering method to discretize the system's phase space, which is not necessarily coupled to the system's energy. In a phase-separating system however, every reasonable discretization scheme that somehow captures the degree of phase-separation will always be, at least loosely, coupled to the system's energy because the degree of phase separation of a configuration is linked to its potential energy (mixed configurations being energetically higher than phase-separated configurations). The 2D-Ising model that has been studied in this work simply constitutes a special case in that the chosen order parameter is lumping together all isoenergetic configurations. MSM is able to extract thermodynamic and dynamic information about these phase-separating systems by gathering information about local fluctuations in energy, i.e. local back- and forth-jumps, between configurations during the (fast) overall evolution from mixed to phase-separated configurations, i.e. from high-energy to low-energy states, and thus avoids the necessity of observing spontaneous unmixing during a simulation to obtain information about the system in equilibrium. This coupling of order parameter and energy in phase-separating systems suggests that the biasing effects that we have observed in the Ising system will also be present in Markov state models of atomistic phase-separating systems.}

The effect of the initial distribution onto the MSM quality and the existence of this bias due to local non-equilibrium has been reported in the literature\cite{Noe2011} and possible approaches to resolve these issues have been proposed\cite{Noe2013-PMM, Noe2015-PMM-OOM}. In particular, these approaches consist of discarding the assumption that dynamics are Markovian on the discretized partition of the phase space, which leads to the concept of Projected Markov Models. Recently, No{\'e} \etal showed that this bias is minimized in the limit of long lag times and fine discretizations and developed a modification of Markov state modeling that exploits observable operator model theory to estimate unbiased MSM transition matrices\cite{Nueske2016}. Simply increasing the lag time in order to reduce this bias is, unlike in other systems such as e.g. proteins, not an option in phase-separating systems. \update{There, an increased lag time would result in a loss of information about the mixed states due to the fast crossing of the high-energy regions and thus lead to failure of the method because the mixed configurations would not be included in the obtained Markov state model.} Whether this new observable operator model based method is able to estimate correct Markov state models of the phase-separating Ising model remains an open question that we will address in future work. In particular, the impact of absent connections between single configurations of Markov states on this bias as well as the question whether this new method is able to resolve this effect remain to be answered.

The Ising model with fixed concentration of up- and down-spins is a prototype for phase-separating systems. It has been used, e.g., to identify generic properties of spinodal decomposition (e.g. by D. A. Huse\cite{Huse1986}, Amar \etal\cite{Amar_etal1988} and Gunton \etal\cite{Gunton_etal1988}). This suggests that the present observations may also be relevant for atomistic phase-separating systems such as lipid membranes. \update{There again, one might choose the number of nearest neighbors of unlike atoms/molecules as an order parameter for the system that captures the degree of phase-separation\cite{Hakobyan2013}. For example, for the standard ternary mixtures of satured lipids, unsaturated lipids and cholesterol one might quantify the free energy gain upon raft formation as a function of composition and temperature.} Of course, in contrast to the present system, the total energy is then only loosely connected with the order parameter due to the large contribution of the surrounding water molecules. Thus, methods like Wang-Landau sampling would be difficult to apply. \update{MSM might be ideal to compute free energies of these systems if the bias due to local non-equilibrium can be quantified and successfully minimized.} 

We would like to stress again that a Markov state model is built from standard molecular dynamics or Monte Carlo data and therefore does not employ external constraints to sample the system's phase space. This is in contrast to other free energy techniques such as, e.g., Umbrella sampling\cite{TorrieValleau1977}, where also thermodynamically unstable states can be sufficiently sampled via some external bias potential. However, preliminary simulations for more complex systems such as lipid membranes indicated that external bias potentials may give rise to a sampling of unphysical configurations. In any event, this has to be explored closer in future work. Thus, the present study constitutes a first step to explore the applicability of MSM to characterize the thermodynamics of unmixing.

%%% --- Supplementary Material: ---
\section*{Supplementary Material}
See supplementary material for further information about the five state model and additional data of the Ising system.

%%% --- Acknowledgements: ---
\begin{acknowledgments}

We gratefully acknowledge helpful discussions with E. Rosta and F. Noe about this work. \update{Furthermore, we thank the \emph{Deutsche Forschungsgemeinschaft} (DFG) for funding via the SFB 858.}

\end{acknowledgments}

%%% --- Appendix: ---
\appendix
\section*{Appendix}
\subsection{Derivation of the analytic solution of the five state model for $\tau>1$}\label{app:Derivation_tau}
First, we have to define the numbers of transitions during one lag time interval (at time $t$): 
\begin{align}
	\overset{\sim}{C}_\mathrm{BB}(t,\tau) &= \left[T_{11}(\tau)+T_{12}(\tau)\right] \times y_1(t-\tau) \notag \\
									      &\,+ \left[T_{21}(\tau)+T_{22}(\tau)\right] \times y_2(t-\tau)\\
	\overset{\sim}{C}_\mathrm{BA}(t,\tau) &= T_{10}(\tau) \times y_1(t-\tau) + T_{20}(\tau) \times y_2(t-\tau) \\
	\overset{\sim}{C}_\mathrm{AA}(t,\tau) &= T_{00}(\tau) \times y_0(t-\tau) \\
	\overset{\sim}{C}_\mathrm{AB}(t,\tau) &= \left[T_{01}(\tau)+T_{02}(\tau)\right] \times y_0(t-\tau) ,
\end{align}
where the $T_{ij}(\tau)$ denote the elements of the respective transition probability matrix for lag time $\tau$, i.e. $\mathbf{T}(\tau) = \left[\mathbf{T}\right]^{\tau}$ with the transition matrix $\mathbf{T}$ from eq. \eqref{eq:pmalMatrix}. The resulting numbers after multiple MC steps, $t$, in the sliding-window mode are then given by
\begin{align}
	C_\mathrm{BB}(t,\tau) &= \sum\limits_{i=\tau}^{t} \overset{\sim}{C}_\mathrm{BB}(i,\tau)
\end{align}
(and analogously for transition counts $C_\mathrm{BA}, C_\mathrm{AB}, C_\mathrm{AA}$).

Based on eq. \eqref{eq:piBpiA}, the estimator $\frac{\pi_\mathrm{B}}{\varepsilon \pi_\mathrm{A}}$ can then be computed from the transition counts via
\begin{align}
	\frac{\pi_\mathrm{B}}{\varepsilon \pi_\mathrm{A}} &= \frac{1}{\varepsilon} \frac{C_\mathrm{AB}C_\mathrm{B}}{C_\mathrm{BA}C_\mathrm{A}} \notag \\
						&= \frac{1}{\varepsilon} \frac{C_\mathrm{AB}(C_\mathrm{BA}+C_\mathrm{BB})}{C_\mathrm{BA}(C_\mathrm{AB}+C_\mathrm{AA})} \notag \\
						&= \frac{1}{\varepsilon} \frac{\left(1+\frac{C_\mathrm{BB}}{C_\mathrm{BA}}\right)}{\left(1+\frac{C_\mathrm{AA}}{C_\mathrm{AB}}\right)}
\end{align}
for various lag times and various numbers of MC steps.

\subsection{Derivation of $t_\mathrm{equil}$, $t_\mathrm{equil,1}$ and the behavior for $t < t_\mathrm{equil,1}$}\label{app:Derivation_tequil}
Here, we start by considering the eigenvalues and eigenvectors of the transition matrix $\mathbf{T}$ in the limit of very small $\varepsilon$. Indeed, these values are well defined in the limit of vanishing $\varepsilon$ and read 
\begin{align}
	\lambda_1 &= 1,\\
	\lambda_{2,3} &= \frac{1}{4} (1 \pm \sqrt{5})
\end{align}
with the corresponding left eigenvectors (before normalization) 
\begin{align}
	\mathbf{ev}_1 &= (1,0,0),\\
	\mathbf{ev}_{2,3} &= \left(\frac{1}{2} ( 3 \pm \sqrt{5}),1,\frac{1}{2} ( 1 \pm \sqrt{5})\right).
\end{align}

In a first step, one can directly read off 
\begin{align}
	t_\mathrm{equil} = -1/\ln(\lambda_2) \approx 5.
\end{align}
This relation is equivalent to the statement that 
\begin{align}
	\lambda_2^{t_\mathrm{equil}} = 1/e.
\end{align}  
Furthermore, since $|\lambda_2| > |\lambda_3|$, there exists a timescale from which on the ratio $y_2(t) / y_1(t)$ is determined by the respective components of the second eigenvector, i.e. 
\begin{align}
	\frac{y_2(t)}{y_1(t)} \approx \frac{1}{2} ( 1 + \sqrt{5}) \approx 1.6.
\end{align}
For $\varepsilon = 0$ this limit would not change any more. For small but finite $\varepsilon$ and for an initial population of Markov state B, this behavior breaks down if $\lambda_2^{t_\mathrm{equil,1}} \approx 4 \varepsilon$ because then the total population of Markov state B approaches its equilibrium value and the presence of upward transitions becomes relevant to establish global equilibrium, in particular $y_1(t) = y_2(t)$. This relation can be rewritten as 
\begin{align}
	t_\mathrm{equil,1} = t_\mathrm{equil} \times \ln (\frac{1}{4\varepsilon}).
\end{align}

%%% --- Bibliography: ---

\end{document}